# A KEY ESTABLISHMENT SCHEME FOR MOBILE WIRELESS SENSOR NETWORKS USING POST-DEPLOYMENT KNOWLEDGE

Ashok Kumar Das

Center for Security, Theory and Algorithmic Research
International Institute of Information Technology, Hyderabad 500 032, India
iitkgp.akdas@gmail.com, ashok.das@iiit.ac.in

## ABSTRACT

*Establishment of pairwise keys between sensor nodes in a sensor network is a difficult problem due to resource limitations of sensor nodes as well as vulnerability to physical captures of sensor nodes by the enemy. Public-key cryptosystems are not much suited for most resource-constrained sensor networks. Recently, elliptic curve cryptographic techniques show that public key cryptosystem is also feasible for resource-constrained sensor networks. However, most researchers accept that the symmetric key cryptosystems are viable options for resource-constrained sensor networks. In this paper, we first develop a basic principle to address the key pre-distribution problem in mobile sensor networks. Then, using this developed basic principle, we propose a scheme which takes the advantage of the post-deployment knowledge. Our scheme is a modified version of the key prioritization technique proposed by Liu and Ning. Our improved scheme provides reasonable network connectivity and security. Moreover, the proposed scheme works for any deployment topology.*

## KEYWORDS

*Security, Key establishment, Mobile sensor networks, Key prioritization, Post-deployment knowledge.*

## 1. INTRODUCTION

Micro Electro-mechanical Systems (MEMS) technology [25] has developed the sensor nodes which are smaller in size, more battery-powered and less production costs. By reducing the size of a sensor node, it is easy to deploy across a target field (i.e., deployment area). The first application of sensor networks was the Sound Surveillance System (SOSUS) [21] which had been used during the cold war in the early 1950s for the purposes to detect as well as track Soviet submarines with the help of acoustic sensors or hydrophones. Due to breakthrough of the MEMS technology, more civilian applications have been observed. There are several applications of sensor networks, for examples, environmental monitoring, classroom/home [9, 24], health monitoring of patients [23], habitat monitoring [3, 18], etc.

In a sensor network, there are many tiny nodes called the sensors, which are deployed for the purposes to sense data and then to transmit the data to the nearby base station. Sensor networks are mainly classified into two categories: distributed and hierarchical. In a *hierarchical wireless sensor network (HWSN)* shown in Figure 1, there is a hierarchy among the nodes based on their capabilities: base stations, cluster heads and sensor nodes. *Sensor nodes* are inexpensive, limited capability and generic wireless devices. Each sensor has limited battery power, memory size and data processing capability and short radio transmission range. Sensor nodes in a cluster (also called a group), communicate each other in that cluster and finally communicate with the cluster head (CH). Usually, *Cluster heads* have more resources than sensors, which are equipped with high power batteries, large memory storages, powerful antenna and data processing capabilities. Cluster heads can execute relatively complicated numerical operations than sensors and have much larger radio transmission range than sensor nodes. Cluster heads can communicate among each other directly and relay data between its cluster members and the base station. A *base station/sink node* (BS) is typically a gateway to another network or an access point for human interface. A base station collects sensor readings, perform costly operations on behalf of sensor nodes and manage the network. In some applications, the base station is assumed to be trusted. Thus, the base station is used as key distribution center (KDC).

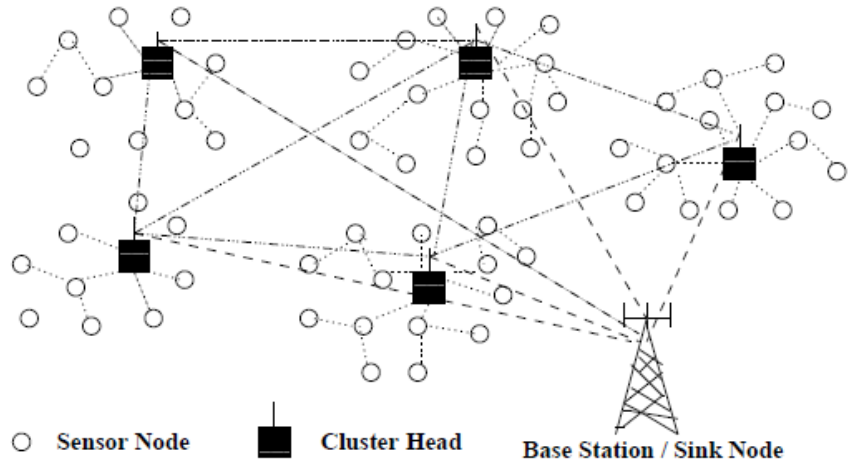

Figure 1: A hierarchical wireless sensor network (HWSN) architecture.

On the other hand, in a *distributed wireless sensor network (DWSN)* shown in Figure 2, there is no fixed infrastructure and network topology is not known prior to deployment of the sensor nodes in the target field. Sensor nodes are usually deployed all over the target area randomly. After deployment, sensor nodes form an infrastructure-less multi-hop wireless communications between them and data is routed back to the base station. Data flow in DWSN is similar to data flow in HWSN with a difference that network-wise (broadcast) flow takes place by every sensor node in the network. In this paper, we consider the distributed sensor networks for developing key establishment scheme.

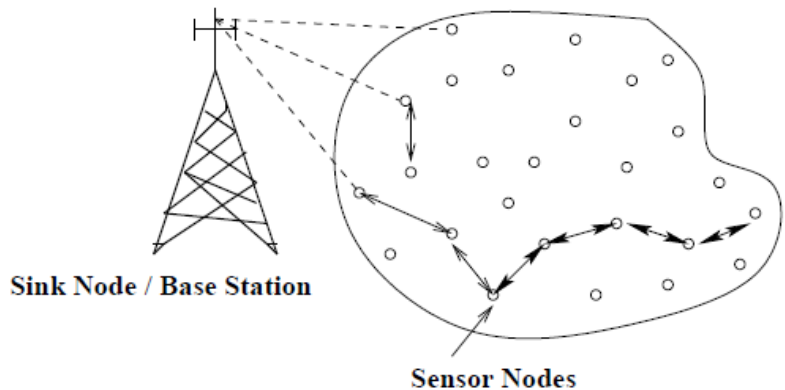

Figure 2: A distributed wireless sensor network (DWSN) architecture.

International Journal of Computer Networks & Communications (IJCNC) Vol.3, No.4, July 2011

The topology of the network is in general dynamic, because the connectivity among the sensor nodes may vary with time due to the factors such as new node addition, node departures, and the possibility of having mobile nodes. In our case, we assume that nodes are mobile and so the network topology is dynamic.

Pairwise key establishment between neighboring sensor nodes in a sensor network is done by using a protocol which is popularly known as the bootstrapping protocol. A bootstrapping protocol usually involves three phases. In key pre-distribution phase, each sensor node is initialized by a set of pre-distributed keys in its memory, called the key ring. This is done before deployment of the sensor nodes by the key setup server (the base station) in a target field. After deployment, a direct key establishment (also called the shared key discovery) phase is performed by sensor nodes in order to establish direct pairwise keys between them. The path key establishment phase is executed if two neighbor nodes fail to establish a direct key after the direct key establishment phase. Suppose that two neighbors $u$ and $v$ fail to establish a direct secure link between them in the direct key establishment phase, but there exists a path, say $\langle u = u_0, u_1, \ldots, u_{h-1}, u_h = v \rangle$ in the network with each $u_i$ being a direct neighbor of $u_{i+1}$ for $i = 0, 1, \ldots, h-1$. $u$ generates a random key $k$, encrypts $k$ with the key shared between $u = u_0$ and $u_1$ and transmits this encrypted key to $u_1$. $u_1$ then retrieves $k$ by decrypting the encrypted key using the shared key between $u$ and $u_1$, encrypts $k$ by the key shared by $u_1$ and $u_2$ and transmits it to $u_2$. This process is continued until the key $k$ reaches to the desired destination $v$. Now $u$ and $v$ can communicate secretly and directly using $k$. The main difficulty in this process is the discovery of a secure path between $u$ and $v$. The communication overhead increases significantly with the number $h$ of hops. Some improvements over the path key establishment phase have been proposed in [26], [27].

A location-adaptive scheme is based on the random deployment models i.e., it does not use any prior knowledge of the deployment locations. Thus, such a scheme has the ability to adjust in any deployment areas. On the other hand, a location-aware scheme depends on the prior knowledge of the deployment locations like the expected location of each sensor and the overall geometry of the deployment area. This pre-deployment knowledge helps us to tune the key pre-distribution algorithms to achieve better network connectivity and higher resilience against node capture. Several techniques [8], [4], [13], [20], [6], [2], [14], [5] are proposed in the literature. All these techniques address only for the static sensor networks. Each and every time when a sensor node moves, it needs to perform the bootstrapping procedure in order to establish direct pairwise keys in that new location. In fact, we must avoid path key establishment phase of the bootstrapping procedure for the mobile sensor networks because it involves much more communication and computational overheads. In this paper, we focus on the issues of the key pre-distributions in mobile sensor networks. We first develop a basic principle for addressing the key pre-distribution problem in mobile sensor networks and finally propose a scheme for mobile sensor networks.

The rest of the paper is organized as follows. Section 2 describes some issues in the mobile sensor networks. In Section 3, we describe briefly the overview of the key prioritization technique using the post-deployment knowledge proposed by Liu and Ning [15]. In Section 4, a basic principle has been proposed to address the key pre-distribution problem in mobile sensor networks. Then, in Section 5, we propose a scheme which follows the basic principle and is based on the modified version of the key prioritization technique. In Section 6, we provide the analysis of our proposed scheme. We provide simulation results of our improved scheme in Section 7. Finally, Section 8 concludes the paper.

## 2. ISSUES IN MOBILE SENSOR NETWORKS

To design a key pre-distribution scheme in mobile sensor networks, we may consider the following issues.

The first issue is that we should not assume any prior knowledge of sensors' locations. However, we can assume the post-deployment knowledge of sensors' locations. This assumption becomes practical due to the following researches. Akyildiz et al. [1] pointed out that "most of the sensing tasks require knowledge of positions" and also "location finding systems are required by many of the proposed sensor network routing protocols". There are several recent advances in determining individual sensor nodes' positions either with a global positioning system (GPS) or local references [12, 19]. Sastry et al. [22], Lazos et al. [11], Du et al. [7] and Liu et al. [16, 17] describe the methods of determining secure locations. Thus, in a mobile sensor network, it is a possible task for sensor nodes to determine their deployment locations securely after deployment. Hence, we can use advantage of post-deployment knowledge in mobile sensor networks.

The second issue is to use extra memory for applications to store an excessive amount of pre-distributed keys as well as the direct pairwise keys between neighbor sensors. Crossbow Technology Inc. [10] develops a typical MICA2 mote sensor device which has 512 EEPROM, but only 4KB RAM. Thus, it is practical to store more pre-distributed keying information in a sensor device.

## 3. KEY PRIORITIZATION TECHNIQUE USING POST-DEPLOYMENT KNOWLEDGE

We describe briefly the concept of the key prioritization technique proposed by Liu and Ning [15]. Their scheme takes the advantage of the post-deployment knowledge of sensor nodes to improve the pairwise key pre-distribution in static sensor networks.

This scheme assigns each sensor node an excessive amount of pre-distributed keys in key pre-distribution phase by using the memory for sensing applications. Then, depending on the post-deployment knowledge, it prioritizes the pre-distributed keys in key prioritization phase, and discard the low priority keys in order to thwart against node capture attack. Since the low priority keys are deleted from the memory, so the returned memory is used for the application part.

In direct key establishment (i.e., shared key discovery) phase, two neighbor nodes establish a pairwise key by exchanging the IDs of the higher priority keys. Liu and Ning then applied it to the polynomial pool-based scheme [13] and its analysis shows that it significantly improves the security and performance than the previous key pre-distribution schemes.

## 4. OUR BASIC PRINCIPLE

In this section, we now introduce a basic principle for key pre-distribution problem in a mobile sensor network.

Let us assume that the deployment area (i.e., the target field) is two-dimensional so that each point in that region can be expressed by two co-ordinates $x$ and $y$. Assume that there are $n$ sensor nodes in a sensor network and each node can hold at most $m$ symmetric cryptographic keys in its memory. Let $K$ be a key pool of size $M$ consisting of key units such that each key unit has a randomly generated key material and a unique location of the deployment field associated with this key material. In other words, $K = \{ku_i = (k_i,(x_i, y_i)) \mid (x_i, y_i)$ is a unique location in the deployment field associated with the key material $k_i\}$, where $ku_i$ is called a key unit in $K$. A key material in the basic probabilistic key pre-distribution scheme (the EG scheme) [8], the q-composite scheme [4], or the random pairwise keys scheme [4] is simply a pre-distributed key. In the polynomial pool-based scheme [13], a key material is a t-degree symmetric bivariate polynomial from which a sensor node can compute keys shared with its neighbors, whereas in the pairwise key pre-distribution scheme [5], a key material is a row of the secret matrix $A_i$ in a key space $S_i$.

We use the following two functions for developing the basic principle as follows:

- $F(\cdot)$ function takes the inputs as (i) $n$, the number of nodes in the network, (ii) $C$, a set of predefined constraints, and (iii) $K$, the key pool. It produces the output as $KR_i$, the sets of $m$ key units called the key rings for the nodes $u_i (i = 1, 2, \ldots, n)$. Thus, a key ring $KR_i$ of a sensor node $u_i$ is $KR_i = \{(k_i,(x_i, y_i)) \mid 1 \leq i \leq m\}$. The (key) setup server generates key rings each of size $m$ for all the sensor nodes to be deployed using the $F(\cdot)$ function. The function $F(\cdot)$ is programmed into the key setup server.
- $G(\cdot)$ function is useful when two neighbors wish to establish a direct pairwise key between them during direct key establishment phase. It takes the inputs as (i) identifiers of nodes, (ii) two public addresses, and (iii) the key ring of a sensor node. Then it produces a secret shared pairwise key between two neighbors. For example, let $u$ and $v$ be two neighbors having the public addresses $PA_u = (u_x, u_y)$ and $PA_v = (v_x, v_y)$ respectively, where $(u_x, u_y)$ and $(v_x, v_y)$ be the locations in the deployment area. Let $ID_u$ and $ID_v$ represent the identifiers of nodes $u$ and $v$, respectively. Then $u$ uses the $G(\cdot)$ function to compute the secret shared key between $u$ and $v$ as $k_{u,v} = G(ID_u, ID_v, PA_u, PA_v, KR_u)$. Similarly, $v$ also uses the same function to compute the secret shared key between $u$ and $v$ as $k_{v,u} = G(ID_u, ID_v, PA_u, PA_v, KR_v)$. Now, if the $G(\cdot)$ function is symmetric one-way function then we have, $k_{u,v} = k_{v,u}$. In this case, nodes $u$ and $v$ need only to exchange their own identifiers and public addresses in order to compute the secret shared key between them. The function $G(\cdot)$ is programmed into each sensor device.

The following basic principle can be applied in mobile sensor networks. This protocol is only based on the post-deployment knowledge and does not assume any pre-deployment knowledge. Our basic protocol consists of the following steps:

- Step 1. After deployment of sensor nodes, the sensor nodes may move from one location to another location in the target field. In the new location, the sensor nodes determine their post-deployment locations. Let $u$ and $v$ be two physical neighbors want to establish a secret pairwise key.
- Step 2. $u$ determines the post-deployment location which is $PA_u = (u_x, u_y)$. $u$ sends its identifier $ID_u$ and $PA_u$ to its neighbor node $v$.

- Step 3. $v$ also determines its post-deployment location $PA_v = (v_x, v_y)$. Then, $v$ sends its own identifier $ID_v$ and this public address $PA_v$ to $u$.
- Step 4. $u$ computes $k_{u,v} = G(ID_u, ID_v, PA_u, PA_v, KR_u)$ as the secret shared key between $u$ and $v$ using the function $G(\cdot)$.
- Step 5. $u$ generates a short puzzle message, say, $P$ and sends the message $(P, E_{k_{u,v}}(P))$ to $v$, where $E$ is a symmetric-key encryption function.
- Step 6. $v$ computes the key $k_{v,u} = G(ID_u, ID_v, PA_u, PA_v, KR_v)$ and decrypts the encrypted puzzle $E_{k_{v,u}}(P)$ using the key $k_{v,u}$. If decryption is successful, that is, if the decrypted puzzle matches with the incoming puzzle, nodes $u$ and $v$ use the secret key $k_{u,v} (= k_{v,u})$ for their future secret communication.

## 5. THE IMPROVED SCHEME

In this section, we propose a key establishment scheme for mobile sensor networks which works for any deployment topology. It follows our basic principle described in Section 4. To achieve this, we use the modified version of the key prioritization technique proposed by Liu and Ning described in Section 3. Thus, our improved scheme is a modified version of the key prioritization technique applied over the basic probabilistic scheme (also known as the EG scheme) [8].

Let $d$ be the average number of the neighbor nodes that each sensor node can contact. Assume that each sensor node will be given a maximum of $m$ pre-distributed keying information in its memory. Each keying information called a key unit, which consists of a symmetric key and a unique location of the deployment area associated with it. As pointed out in Section 2, since we can use memory from application in order to store an excessive amount of keying materials, so we can use memory for storing $m + d$ keying information in each sensor device. The low priority key units from the per-loaded $m$ key units will not be deleted from the memory even after the key prioritization phase. We reserve memory for updating direct keys established between a sensor node and its neighbor nodes due to change in the network topology such as the mobility of sensor nodes and addition of new sensor nodes.

The different phases of our scheme are as follows.

### 5.1. Key pre-distribution phase

In this phase, the key setup server randomly generates a set $K$ of $M$ key units, called the key pool such that each key unit is a pair $(k, (x, y))$ where $k$ is a random number generated by the key setup server and $(x, y)$ is a unique deployment location associated with $k$. The key setup server also assigns a unique ID to each sensor node. Let there be $n$ sensor nodes in the network and $ID_{u_i}$ be the identifier of a sensor node $u_i (i = 1, 2, \ldots, n)$. For each sensor node $u_i$, the key setup server picks a subset of $m$ key units randomly from the key pool $K$ and then distributes this subset to that sensor node before its deployment in the target field.

## 5.2. Key prioritization phase

Immediately after deployment, the sensor nodes will determine their post-deployment locations. Two nodes are called *physical neighbors* if they are within the communication ranges of one another. Let $u$ and $v$ be two physical neighbors. The public addresses of $u$ and $v$ are considered as their post-deployment locations, that is, $PA_u = (u_x, u_y)$ and $PA_v = (v_x, v_y)$. Then, each sensor node computes the distances between its post-deployment location and the locations associated with the pre-distributed key units. The smaller distance has the higher priority. The sensor nodes then choose $c$ high priority key units and keep the remaining low priority key units in its key ring. As a result, the sensor nodes close to each other are more likely to have a higher probability to establish a common key.

## 5.3. Direct key establishment phase

Two nodes are called *key neighbors* if they share a common key in their key rings. They are called *direct neighbors* if they are both physical neighbors as well as key neighbors. Now, to establish a direct key between two neighbor nodes, they only need to identify a common key shared between them. This is achieved by exchanging the locations of $c$ high priority key units between them. Let two nodes $u$ and $v$ share $q\ (\geq 1)$ common keys, say, $k_1, k_2, \ldots, k_q$. The secret key $k_{u,v}$ shared between nodes $u$ and $v$ is computed by each other as $k_{u,v} = h(ID_u \parallel ID_v \parallel (PA_u.x + PA_v.x) \parallel (PA_u.y + PA_v.y) \parallel k_1 \parallel k_2 \parallel \ldots \parallel k_q)$, where $h(\cdot)$ is a one-way hash function (for example, SHA-1), $PA_u.x$ and $PA_u.y$ the x-coordinate and y-coordinate of the post-deployment location $PA_u$ of a sensor node $u$, and $\parallel$ the concatenation operation. Then nodes can verify their established secret key $k_{u,v}$ using the challenge-response protocol described in Steps 5 and 6 of Section 4.

## 5.4. Dynamic node addition phase

The addition of a new sensor node is simple. The key setup server needs to perform the above key pre-distribution phase in order to load $m$ key units in its memory. The key setup server also assigns a unique identifier to that node. After deployment in the field, the key prioritization and direct key establishment phases will be performed in order to establish secret keys with the neighbor nodes by the newly deployed node.

## 6. ANALYSIS OF THE IMPROVED SCHEME

In this section, we analyze the network connectivity and security against node capture with respect to our improved scheme.

### 6.1. Probability of establishing direct keys between neighbors

Assume that $u$ and $v$ are two neighbor nodes such that their distance is $x$. Let $m$ be the key ring size of each sensor node, $d$ the average number of nodes that a sensor node can contact, and $d_r$ be the communication range of each sensor node. We take $d_r$ as the basic unit of measure distance $d_r = 1$.

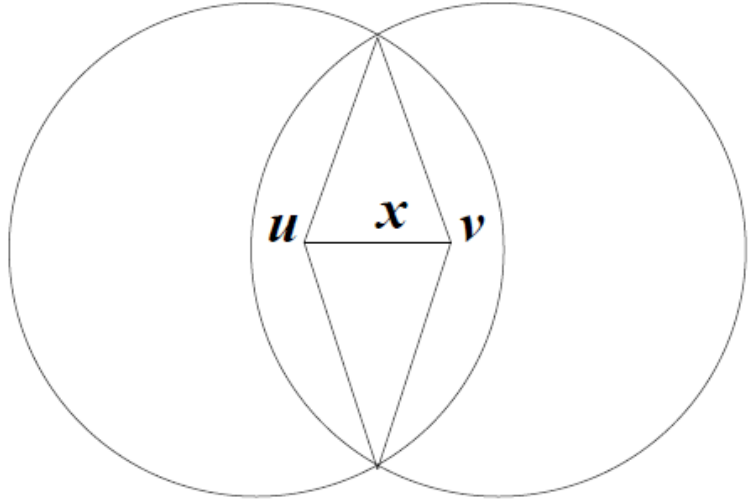

Figure 3: Overlapping area between two neighbor nodes $u$ and $v$.

If $A$ is the size of the deployment field, the maximum number of nodes in the network becomes $n = \left\lfloor \dfrac{A \times (d+1)}{\pi d_r^2} \right\rfloor = \left\lfloor \dfrac{A \times (d+1)}{\pi} \right\rfloor$. Since a node $u$ gets $m$ keys during the key pre-distribution phase and ranks only $c$ high priority keys after the key prioritization phase, so these $c$ keys' associated locations are no more than a distance $r'$ away from $u$, where $r' = \sqrt{\dfrac{A \times c}{\pi \times m}} = \sqrt{\dfrac{c \times n}{m \times (d+1)}}$. The overlap area within both nodes' (i.e., $u$ and $v$) communication radii (shown in Figure 3) is given by

$$A(x) = 2r'^2 \cos^{-1}\left(\dfrac{x}{2r'}\right) - x\sqrt{r'^2 - \dfrac{x^2}{4}}.$$

The (average) number of pre-distributed keys that fall into $A(x)$ is given by

$$N'(x) = \left\lfloor m \times \dfrac{A(x)}{A} \right\rfloor, \text{ and}$$

the total number of keys that are distributed over $A(x)$ is

$$N'(x) = \left\lfloor M \times \dfrac{A(x)}{A} \right\rfloor,$$

where $M$ is the size of the key pool. Thus, the probability that two neighbor nodes $u$ and $v$ share a key is computed as

$$p(x) = 1 - \frac{\binom{N(x)-N'(x)}{N'(x)}}{\binom{N(x)}{N'(x)}} = 1 - \prod_{i=0}^{N'(x)-1} \frac{N(x)-N'(x)-i}{N(x)-i}.$$

Hence, the (average) probability of establishing a direct key between two neighbor nodes is given by

$$p = \int_{r=0}^{d_r} \int_{\theta=0}^{2\pi} \frac{p(r)}{\pi d_r^2} r\, dr\, d\theta = 1 - 2\int_0^1 f(r)\, dr,$$

where $f(r) = r \times \prod_{i=0}^{N'(r)-1} \frac{N(r)-N'(r)-i}{N(r)-i}.$

Figure 4 illustrates the relationship between the average number of nodes $d$ of a sensor node and the maximum supported network size $n$ for our proposed scheme. It is clear from this figure that the maximum supported network size increases as the average number of nodes of a sensor node also increases. Moreover, it is also easy to observe that the maximum supported network size decreases as the communication range $d_r$ increases.

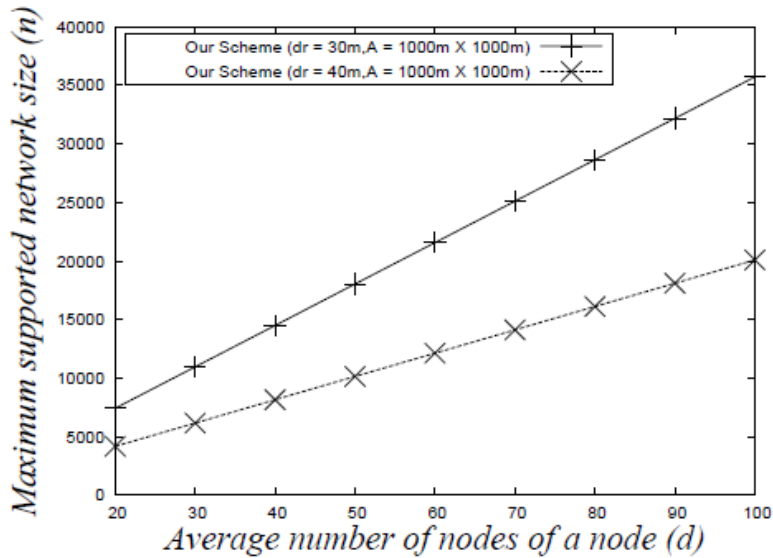

Figure 4: The average number of nodes $d$ of a sensor node *versus* the maximum supported network size $n$. Assume that the communication range $d_r$ = 30m, 40m, and the size of the deployment field A = 1000m × 1000m.

In order to compute the probability $p$ of establishing a pairwise key between two neighbor nodes, we use the following parameters:

- The number of sensor nodes in the sensor network is $n = 18000$.
- The size of the deployment field is $A = 1000m \times 1000m$.
- The communication range is $d_r = 40\,m$.

International Journal of Computer Networks & Communications (IJCNC) Vol.3, No.4, July 2011

- The average number of neighbors of a sensor is $d = 50$.
- The key pool size is $M = 100000$.
- The number of pre-distributed keys to each sensor node is $m = 200$.
- The average number of high priority keys after the key prioritization phase is $c = 200$.

We have used the Simpson's $1/3$ rule of integration for an approximate solution of $p$. The direct network connectivity $p$ is shown in Figure 5. When no additional memory is allocated to store pre-distributed keys in the key pre-distribution phase, the probability of sharing a direct key between neighbor nodes is around $0.32$. It is clear from this figure that the network connectivity increases as the percentage of additional memory also increases.

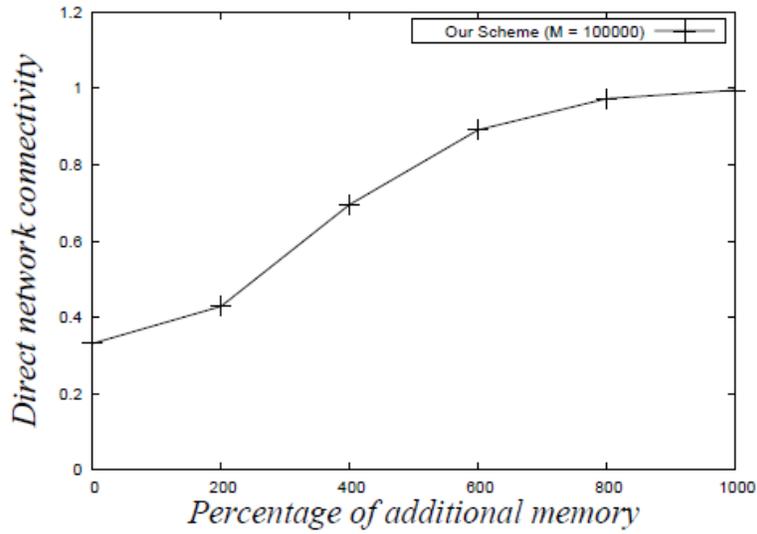

Figure 5: Probability of sharing a direct key between neighbor sensor nodes. Assume $M = 100000, c = 200$ so that $p \approx 0.32$.

### 6.2. Resilience against node capture attack

The resilience against node capture attack of a key distribution scheme is measured by estimating the fraction of total secure communications that are compromised by a capture of $N_c$ nodes *not including* the communication in which the compromised nodes are directly involved. In other words, we want to find out the probability that the adversary can decrypt the secret communications between two non-compromised nodes $u$ and $v$ when $N_c$ sensor nodes are already compromised.

The resilience against node capture of our improved scheme remains same as that of the EG scheme [8] and is given by

$$P_e(N_c) = 1 - \left(1 - \frac{m}{M}\right)^{N_c},$$

where $m$ is the key ring size of a node, $M$ the key pool size, and $N_c$ the number of already captured nodes by an attacker.

The resilience against node capture attack of our improved scheme is shown in Figure 6. It increases as the number of captured nodes also increases. We note that if the sensor nodes are compromised before the key prioritization phase, they may reveal more secrets than compromising the same set of sensor nodes after the key prioritization phase. Thus, an adversary can observe this property and attack the network in the time period between the key pre-distribution and the key prioritization phases. This time period is known as the *window of vulnerability*. However, once a sensor node securely determines its post-deployment location after its deployment in the target field, it can immediately complete the key prioritization. In case of the mobile nodes, the network topology is dynamic and the network connectivity may vary at any time. Hence, it is desirable to assume that nodes are not compromised during the short period of window of vulnerability. Again, when the topology of the network changes due to mobility of sensor nodes, the nodes establish different keys with their neighbors using the deployment location of the target field, the high priority keys and the identifiers. Hence, each and every time when topology of the network changes, the attacker needs to compute the secret keys between non-compromised nodes using the keying information from captured nodes.

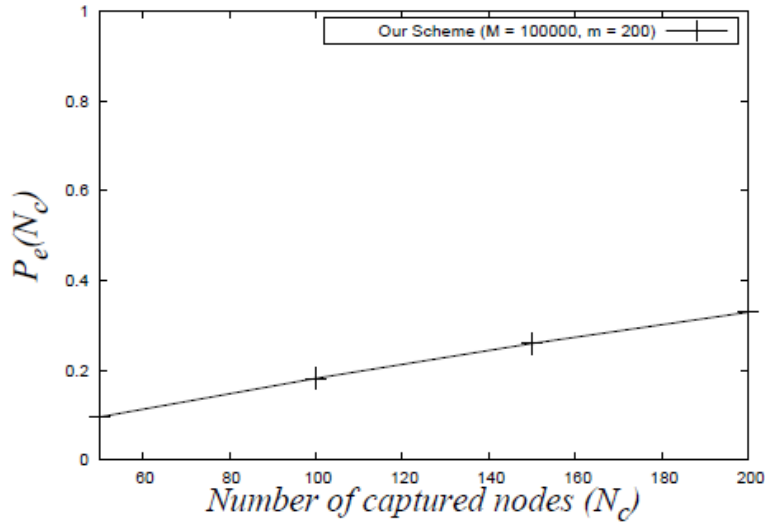

Figure 6: Probability of compromising a secure link between two non-compromised nodes $P_e(N_c)$ *versus* the number of captured nodes, $N_c$. Assume $M = 100000, m = 200, c = 200$ so that $p \approx 0.32$.

## 7. SIMULATION RESULTS

In this section, we have simulated the network connectivity of our improved scheme. For the sake of simplicity, we have taken a square target field. We assume that the deployment field is two-dimensional, so that every point in the region is expressed by two co-ordinates $x$ and $y$. A point $(x, y)$ in the deployment field is considered as a deployment location for a sensor node. Nodes are deployed randomly in the deployment field and due to this, the sensors deployment locations are taken as random points in the deployment field. Finally, we take the deployment location of each sensor node as the post-deployment location of that node.

We have considered the following parameters for the simulation:

- The number of sensor nodes in the sensor network is $n = 18000$.
- The size of the deployment field is $A = 1000m \times 1000m$.

- The communication range is $d_r = 40\,\text{m}$.
- The average number of neighbors of a sensor is $d = 50$.
- The key pool size is $M = 100000$.
- The number of pre-distributed keys to each sensor node is $m = 200$.
- The average number of high priority keys after the key prioritization phase is $c = 200$.

The simulation results for network connectivity versus the analytical results for network connectivity are shown in Figure 7. It is noted that both results tally closely. When no additional memory is allocated to store pre-distributed keys in the key pre-distribution phase, the probability of sharing a direct key between neighbor nodes is around $0.32$. The network connectivity also increases as the percentage of additional memory increases.

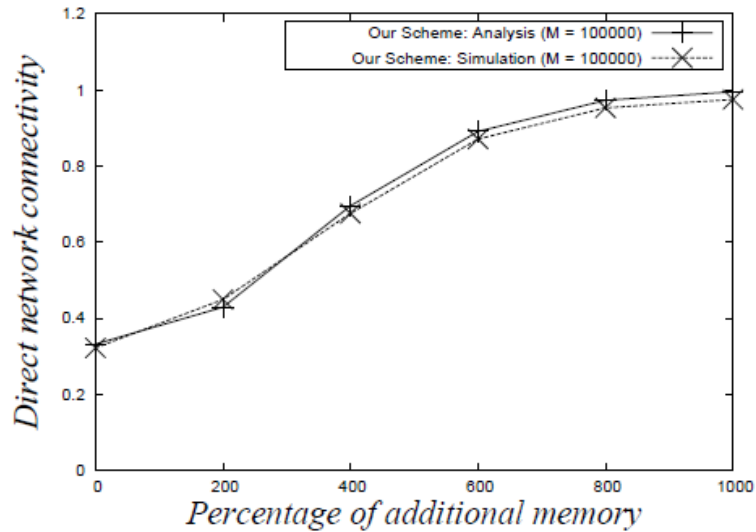

Figure 7: Probability of sharing a direct key between neighbor sensor nodes. Assume $M = 100000$ and $c = 200$.

## 8. CONCLUSIONS

We have proposed a basic principle which gives a general framework to address the key pre-distribution issues in a mobile sensor network. After that, we have proposed a scheme in mobile sensor networks which is based on this basic principle and the modified version of the key prioritization technique. The improved scheme makes use of post-deployment locations of sensor nodes to facilitate the key establishment between neighbor nodes. We have shown that our proposed scheme provides reasonable network connectivity as well as resilience against node compromise. In future, we would like to further improve the resilience against node capture of the proposed scheme. The improved scheme works for any deployment topology and also supports the addition of new sensor nodes after initial deployment.

### ACKNOWLEDGMENT

The author would like to thank the anonymous reviewers for their useful suggestions and comments for improving the paper.

**Author**

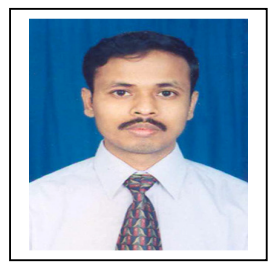

**Dr. Ashok Kumar Das** is currently working as an Assistant Professor in the Center for Security, Theory and Algorithmic Research of the International Institute of Information Technology (IIIT), Hyderabad 500 032, India. Prior to joining IIIT Hyderabad, he held academic position as an Assistant Professor in Department of Computer Science and Engineering of the International Institute of Information Technology, Bhubaneswar 751 013, India from July 2008 to May 2010. He received his Ph.D. degree in Computer Science and Engineering from the Indian Institute of Technology, Kharagpur, India on April 2009. He received the M.Tech. degree in Computer Science and Data Processing from the Indian Institute of Technology, Kharagpur, India on January 2000. He also received the M.Sc. degree in Mathematics from the Indian Institute of Technology, Kharagpur, India, in 1998. Prior to join in Ph.D., he worked with C-DoT (Centre for Development of Telematics), a premier telecom technology centre of Govt. of India at New Delhi, India from March 2000 to January 2004. His current research interests include cryptography, security in wireless sensor networks, mobile adhoc networks and vehicular adhoc networks, proxy ring signature, remote user authentication using smart cards and hierarchical access control. He has published over 20 papers in international journals and conferences in these areas.